\documentclass[aps,prb,preprint,nopacs,superscriptaddress,]{revtex4}

\usepackage{graphicx}
\usepackage{verbatim}
\usepackage{mathrsfs}
\usepackage{gensymb}
\usepackage{color}
\usepackage{ulem}
\usepackage{times}

\usepackage{etoolbox}
\pretocmd{\abstractname}{\newpage}{}{}

\usepackage{amsmath,amsfonts,amssymb}
\usepackage{graphicx}

\newcommand{\bee}{\begin{equation}}
\newcommand{\ee}{\end{equation}}

\def\3{2.5in}    %used for figure widths
\def\2{2.5in}
\def\4{3.0in}\def \beq {\begin{equation}}
\def \eeq {\end{equation}}
\begin{document}

\title{Observation of Weyl fermions in a magnetic non-centrosymmetric crystal}

\author{Daniel S. Sanchez $^*$\footnote[0]{*These authors contributed equally to this work}}
\affiliation {Laboratory for Topological Quantum Matter and Advanced Spectroscopy (B7), Department of Physics, Princeton University, Princeton, New Jersey 08544, USA} 

\author{Guoqing Chang $^*$} 
\affiliation {Laboratory for Topological Quantum Matter and Advanced Spectroscopy (B7), Department of Physics, Princeton University, Princeton, New Jersey 08544, USA}

\author{Ilya Belopolski$^*$} 
\affiliation {Laboratory for Topological Quantum Matter and Advanced Spectroscopy (B7), Department of Physics, Princeton University, Princeton, New Jersey 08544, USA}

\author{Hong Lu}
\affiliation{International Center for Quantum Materials, School of Physics, Peking University, China}

\author{Jia-Xin Yin}
\affiliation {Laboratory for Topological Quantum Matter and Advanced Spectroscopy (B7), Department of Physics, Princeton University, Princeton, New Jersey 08544, USA}

\author{Nasser Alidoust}
\affiliation {Laboratory for Topological Quantum Matter and Advanced Spectroscopy (B7), Department of Physics, Princeton University, Princeton, New Jersey 08544, USA}\affiliation{Rigetti Computing, Berkeley, California 94720, USA}

\author{Xitong Xu}
\affiliation{International Center for Quantum Materials, School of Physics, Peking University, China}

\author{Tyler A. Cochran}
\affiliation {Laboratory for Topological Quantum Matter and Advanced Spectroscopy (B7), Department of Physics, Princeton University, Princeton, New Jersey 08544, USA}

\author{Xiao Zhang}
\affiliation{International Center for Quantum Materials, School of Physics, Peking University, China}

\author{Yi Bian}
\affiliation{International Center for Quantum Materials, School of Physics, Peking University, China}

\author{Songtian S. Zhang}
\affiliation {Laboratory for Topological Quantum Matter and Advanced Spectroscopy (B7), Department of Physics, Princeton University, Princeton, New Jersey 08544, USA}

\author{Yi-Yuan Liu}
\affiliation{International Center for Quantum Materials, School of Physics, Peking University, China}

\author{Jie Ma}
\affiliation{Key Laboratory of Artificial Structures and Quantum Control, School of Physics and Astronomy, Shanghai Jiao Tong University, China}

\author{Guang Bian}
\affiliation{Department of Physics and Astronomy, University of Missouri, Columbia, MO, USA}

\author{Hsin Lin}
\affiliation{Institute of Physics, Academia Sinica, Taipei 11529, Taiwan}

\author{Su-Yang Xu}
\affiliation {Laboratory for Topological Quantum Matter and Advanced Spectroscopy (B7), Department of Physics, Princeton University, Princeton, New Jersey 08544, USA}

\author{Shuang Jia}
\affiliation{International Center for Quantum Materials, School of Physics, Peking University, China}\affiliation{Collaborative Innovation Center of Quantum Matter, Beijing,100871, China}

\author{M. Zahid Hasan $^\dagger$\footnote[0]{$^\dagger$ Corresponding Author: mzhasan@princeton.edu}}
\affiliation{Laboratory for Topological Quantum Matter and Advanced Spectroscopy (B7), Department of Physics, Princeton University, Princeton, New Jersey 08544, USA}\affiliation{Princeton Institute for Science and Technology of Materials, Princeton University, Princeton, New Jersey 08544, USA}
\affiliation{Lawrence Berkeley National Laboratory, Berkeley, CA 94720, USA}

\maketitle

\textbf{Characterized by the absence of inversion symmetry, non-centrosymmetric materials are of great interest because they exhibit ferroelectricity, second harmonic generation, emergent Weyl fermions  and other fascinating phenomena. It is expected that if time-reversal symmetry is also broken, additional magneto-electric effects can emerge from the interplay between magnetism and electronic order. Here we report topological conducting properties in the non-centrosymmetric magnet PrAlGe. By photoemission spectroscopy, we observe an arc parametrizing surface-localized states---a topological arc. Using the bulk-boundary correspondence, we conclude that these arcs correspond to projected topological charges of $\pm{1}$ in the surface Brillouin zone, demonstrating the presence of magnetic Weyl quasiparticles in bulk. We further observe a large anomalous Hall response, arising from diverging bulk Berry curvature fields associated with the magnetic Weyl band structure. Our results demonstrate a topological phase with robust electronic surface states and anomalous transport in a non-centrosymmetric magnet for the first time, providing a novel material platform to study the interplay between magnetic order, band topology and transport.}

%Introduction
The arrangement of atoms in a material determines the symmetry constraints imposed on its electron wavefunctions. A refined understanding of symmetry considerations and their relationship to the band structure Berry curvature has unveiled new and unusual phenomena that can be explored \cite{Anderson,add0,add1,add2,UniverseinHe,revadd,rev2,rev3,rev4,Topo.gapless.phase,Wan,Weyl-Multilayer,HgCr2Se4,RAlGe_Theory,q.spin_current,BerryWavefucn,add3,SHG}. A well-recognized example includes the intrinsic anomalous Hall response arising from Berry curvature in systems breaking time-reversal symmetry. Materials which are non-centrosymmetric, meaning that the crystal structure breaks inversion symmetry, give rise to other exotic phenomena: non-local gyrotropic effects \cite{gyrotropy}, quantum nonlinear Hall effects \cite{q.nonlinearHall}, photogalvanic effects \cite{CPGE}, accidental two-fold band degeneracies (Weyl fermions) that are protected by a quantized non-zero integer Chern number (chiral charge) \cite{ARPES-TaAs1,  ARPES-TaAs3,add4,add5}, and anomalous transport \cite{Chiral-anomaly}. Here, we discover a new non-centrosymmetric magnet, PrAlGe \cite{RAlGe_Theory}, and search for emergent topological properties of magnetic Weyl fermions by spectroscopy, transport and \textit{ab initio} calculation. In contrast to previous works on magnetic Weyl semimetal candidates Mn$_3$Sn \cite{add6,C.MagneticWSM} and Co$_3$Sn$_2$S$_2$ \cite{ Felser_CoSnS,add7,add8} (both centrosymmetric), PrAlGe is calculated to exhibit Weyl fermions in proximity to the Fermi level, making it more suitable for experimentally probing its Berry curvature properties and exploring the connection between photoemission-based band structure and transport. Moreover, because PrAlGe lacks both inversion and time-reversal symmetry (non-centrosymmetric and magnetic) it can uniquely induce quantum spin currents without a concomitant charge current \cite{q.spin_current, RAlGe_Theory}. Motivating future studies on PrAlGe, we experimentally resolve key topological signatures of magnetic Weyl fermions by relying predominately on our measurements \cite{Wan, arcDetect1}.

\begin{figure}
\centering
\includegraphics[width=165mm]{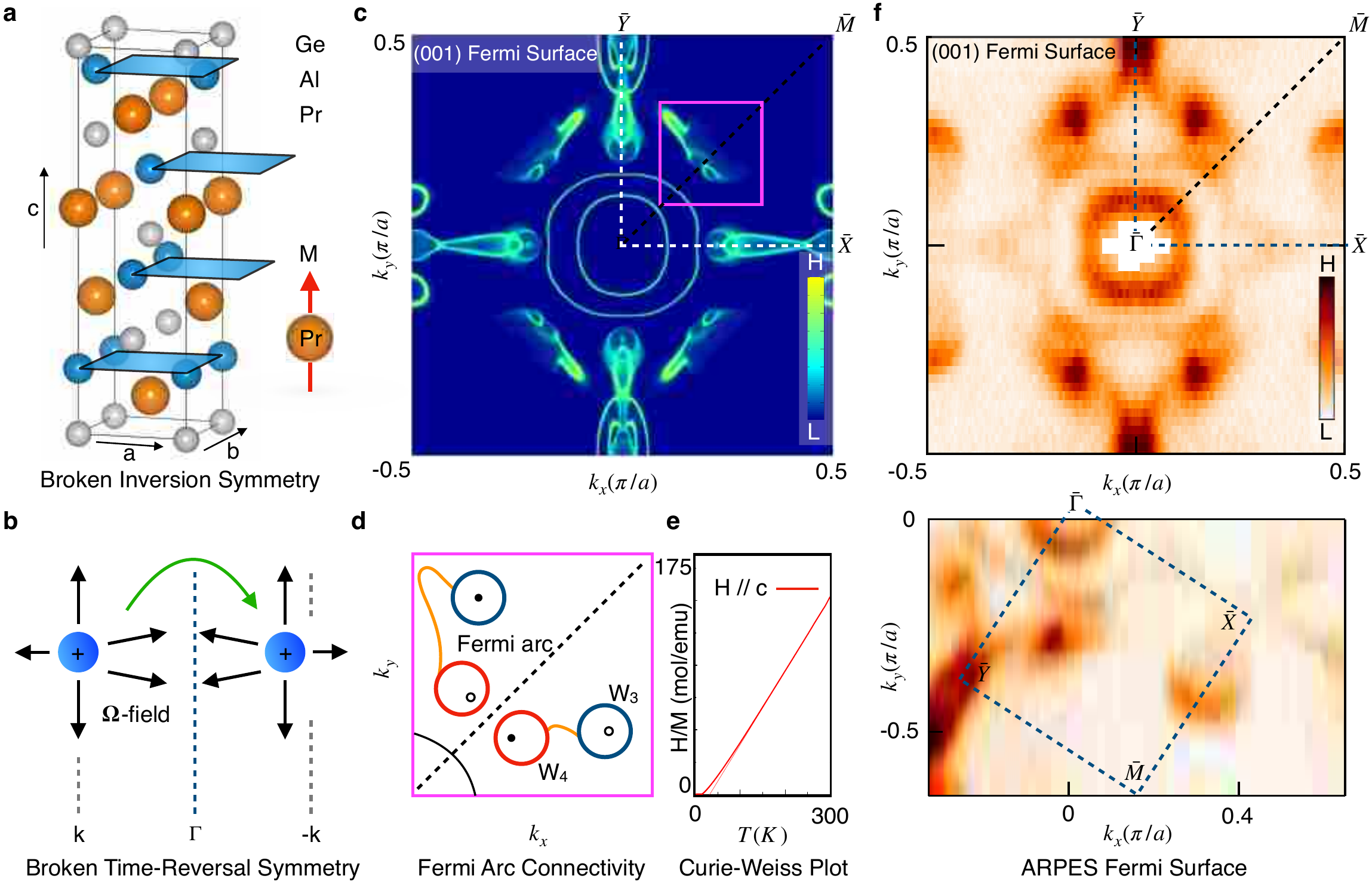}
\caption{\textbf{Lattice and electronic structure of non-centrosymmetric ferromagnet PrAlGe.} 
    \textbf{a}, Crystal structure of PrAlGe in space group $I4_{1}md$ (No. 109). The lattice is built up as a stack of square lattice layers of Pr, Al or Ge. The stacking pattern results in broken inversion symmetry. \textbf{b}, Illustration of the magnetic Weyl semimetal phase of PrAlGe (broken crystal inversion and time-reversal symmetries). Breaking time-reversal symmetry allows the Weyl fermions ($+$), represented as sources of Berry curvature $\Omega$ fields, to be shifted in the crystal momentum space so they are no longer appear pairwise at $\pm{}\textbf{\textit{k}}$. \textbf{c}, \textit{Ab initio} calculated band structure for the (001) surface. White dashed box: first quadrant of the surface BZ. Magenta box: region containing the projected W$_3$ and W$_4$ Weyl fermions, connected by two Fermi arcs. \textbf{d}, Cartoon of the Fermi arc connectivity (orange lines) for the Weyl fermions (black and white circles), corresponding to the magenta box in (c). The predicted configuration of Weyl points manifestly breaks time-reversal symmetry. \textbf{e}, Inverse magnetic susceptibility as a function of temperature (thick line) with fit to a Curie-Weiss law (thin line). The fit gives a positive Weiss constant, indicating the presence of ferromagnetic interactions. \textbf{f}, Top: symmetrized and, bottom: raw, un-symmetrized Fermi surface obtained by ARPES.}
\label{Fig1} 
\end{figure}

%Figure1
PrAlGe crystallizes in a body-centered tetragonal Bravais lattice with space group $I4_{1}md$ (No. 109). The basis consists of two Pr, two Al and two Ge atoms, Fig.~\ref{Fig1}\textbf{a}. Along the (001) direction, each atomic layer is comprised of one element, and the layer is shifted relative to the one below by half a lattice constant in either the $x$ or $y$ direction. This results in a screw pattern along the $z$ direction, associated with a non-symmorphic rotation symmetry that consists of $C_4$ followed by a $z$ translation by $c/4$. Single crystal X-ray diffraction suggests that our samples possess the correct lattice structure and lack inversion symmetry. PrAlGe was predicted to host a ferromagnetic phase with emergent Weyl fermions in $\textit{ab initio}$ calculations\cite{RAlGe_Theory}. The distribution of Weyl fermions in the Brillouin zone (BZ) can be understood as being dominated by the non-centrosymmetric crystal structure with additional Weyl fermions and a momentum-space shift of the Weyl fermions associated with broken time-reversal symmetry, Fig.~\ref{Fig1}\textbf{b}. In particular, an $\textit{ab initio}$ calculation found that certain Weyl fermions, labelled W$_4$, to be discussed below, appeared only when magnetic interactions were taken into account\cite{RAlGe_Theory}. In addition, from basic theoretical considerations, the time-reversal breaking momentum-space shift allows the Weyl fermions to produce an anomalous Hall response. The $\textit{ab initio}$ calculated Fermi surface of PrAlGe further reflects the $\mathcal{C}_{2z}$ symmetry preserved on the (001) surface and when the magnetization is non-zero\cite{RAlGe_Theory}, Fig.~\ref{Fig1}\textbf{c}. Calculation also predicts that within approximately $\pm20$ meV of the Fermi level, only Weyl fermions W$_3$ and W$_4$, as labelled in Ref. \cite{RAlGe_Theory}, are relevant to the low-energy physics, Fig.~\ref{Fig1}\textbf{d}. Moreover, on the (001) surface, topological arcs are predicted to connect the surface projection of these W$_3$ and W$_4$ Weyl fermions. To explore the predicted ferromagnetic state, we first measure the magnetic susceptibility of our single crystal PrAlGe samples as a function of temperature, Fig.~\ref{Fig1}\textbf{e}. We fit the inverse susceptibility to a Curie-Weiss law and obtain a positive Weiss constant, indicating the presence of ferromagnetic interactions. A direct measurement, to be discussed below, shows that PrAlGe is ferromagnetic with Curie temperature $T_\textrm{C} = 16$ K.

\begin{figure}[t]
\centering
\includegraphics[width=165mm]{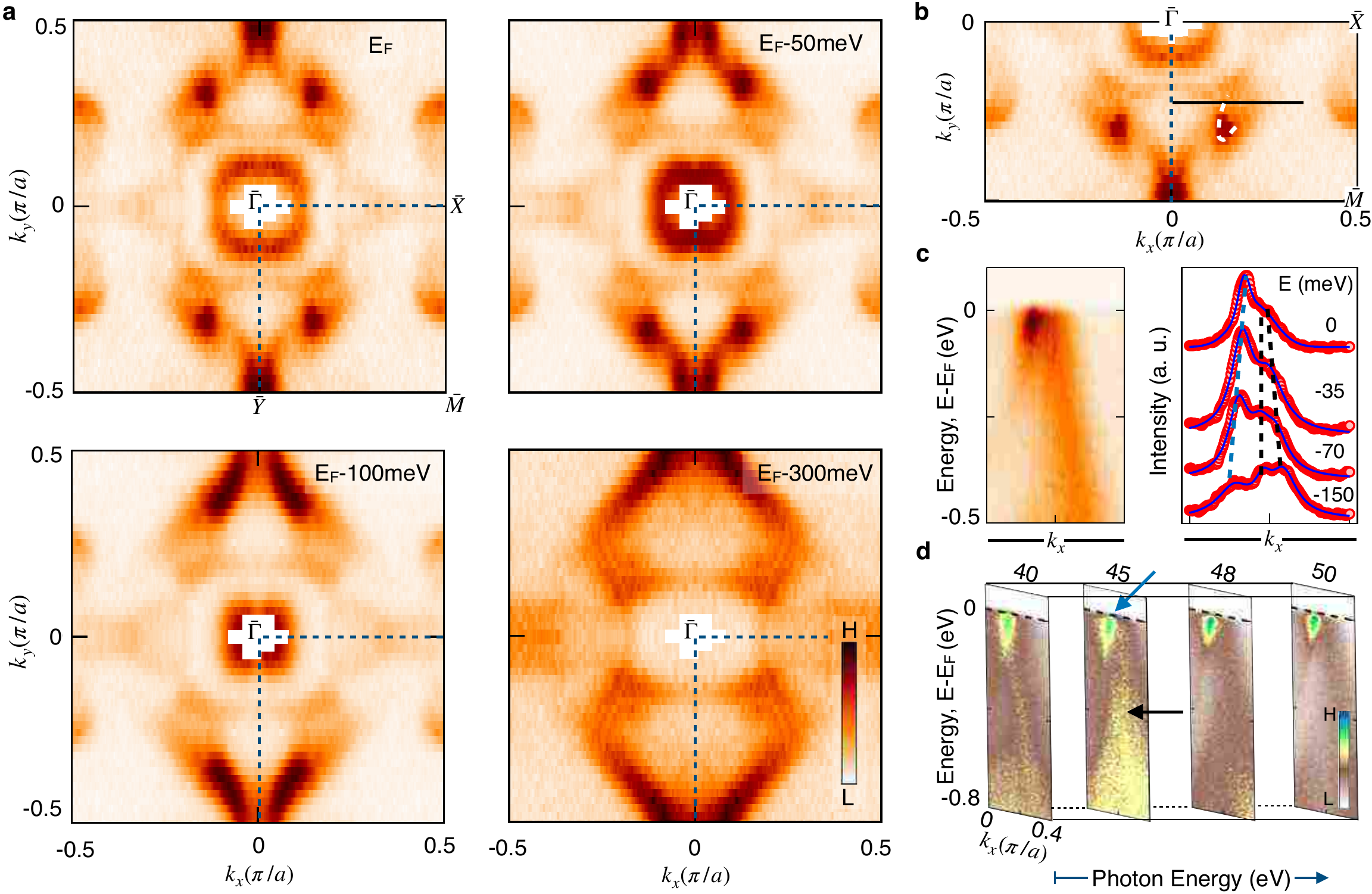}
\caption{\textbf{Fermiology and topology of the (001) surface electronic structure in PrAlGe.} \textbf{a}, ARPES-measured Fermi surface and constant binding energy contours obtained with an incident photon energy of 50eV at $T\sim 11$K. Blue dashed line: one quadrant of the surface Brillouin zone. \textbf{b}, ARPES-measured Fermi surface with a guides to the eye (white dashed line) tracking the `U' shaped candidate topological arc state. \textbf{c}, Left: energy-momentum cut and, right: MDCs fitted at different binding energies with Lorentzian functions to track the candidate arc (blue dashed line) and bulk states (black dashed lines). The corresponding path is shown in (b). \textbf{d}, Photon-energy dependent ARPES along the horizontal line in (b). No $k_z$ dispersion is observed for the candidate `U' shaped topological arc (blue arrow). Strong photon-energy dependence is observed for other states nearby (black arrow), suggesting that they are bulk states.}
\label{Fig2}
\end{figure}

%Figure 2
Motivated by the theoretical prediction\cite{RAlGe_Theory} and our transport measurement discovering ferromagnetism, we use angle-resolved photoemission spectroscopy (ARPES) to map the band structure of PrAlGe on the (001) surface at temperature $T = 11$ K, below $T_c$. We observe that the \textit{ab initio} calculation is qualitatively consistent with the measured Fermi surface, Fig.~\ref{Fig1}\textbf{f}. On constant-energy contours of varying binding energy we observe the following dominant features, Fig.~\ref{Fig2}\textbf{a}: two concentric closed contours around the $\bar{\Gamma}$ point, a distinct `U' shaped state (marked by a guide to the eye, Fig.~\ref{Fig2}\textbf{b}), and additional states near the edge of the surface BZ. The inner closed concentric contour shrinks with deeper binding energy, showing a clear electron-like behavior. To better understand the nature of the `U' state and the spectral intensity in its vicinity, we study an energy-momentum cut through this state, Fig.~\ref{Fig2}\textbf{c}. We plot the Lorentzian-fitted momentum distribution curves (MDCs) at different binding energies and find that the `U' state disperses towards the Fermi level while a nearby band approaches $E_\textrm{F}$ and then turns back towards deeper binding energies. We also find that the `U' state exhibits negligible photon energy dependence, suggesting that it is a surface state (Fig.~\ref{Fig2}\textbf{d}). A comparison with the \textit{ab initio} calculated Fermi surface further suggests that the `U' state corresponds to the predicted Fermi arc connecting W$_3$ and W$_4$. The surface state nature of the `U' state and its correspondence with calculation suggests that the ARPES-measured state is a Fermi arc.

% Focusing on two neighboring quadrants, we mark the $\textbf{\textit{k}}$ space trajectory of the `U' state with a guide to the eye, Fig.~\ref{Fig2}\textbf{b}.

\begin{figure}[t]
\includegraphics[width=165mm]{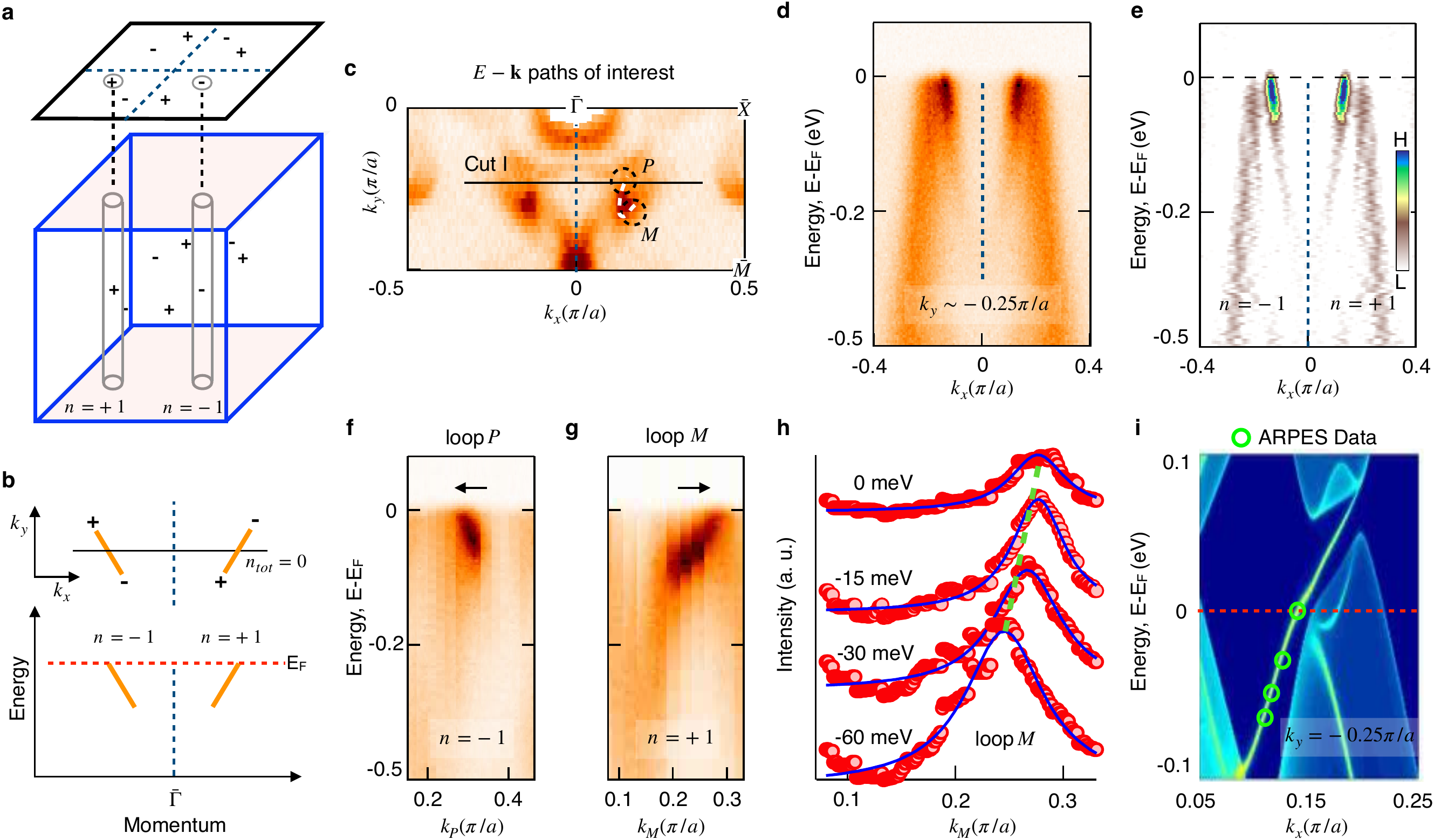}
\caption{ \textbf{Topological arcs and chiral charges in PrAlGe.} 
\textbf{a}, Bulk and surface Brillouin zone (BZ) with the Weyl fermions ($\pm{}$) and manifolds with Chern number $n$ \cite{arcDetect1}. A closed loop enclosing the projected chiral charge in the surface corresponds to a cylinder in the bulk enclosing the Weyl fermion. \textbf{b}, Top: topological arcs (orange) connecting the projected Weyl fermions carrying chiral charge $\pm{1}$. Bottom: a cut across two arcs (along the black line in top panel) with chiral edge modes (orange lines). The two arcs are related by a mirror symmetry (blue dashed line) and propagate in opposite directions, corresponding to a net Chern number of $n_{tot}=0$. \textbf{c}, Fermi surface obtained by ARPES at $T \sim 11$ K. \textbf{d}, Measured band dispersion along horizontal Cut I ($\textit{k}_{y}\sim-0.25\pi/a$), as marked in (c). Blue dashed line: indicates the mirror symmetry. \textbf{e}, Second-derivative plot of (d). Chiral modes with opposite Fermi velocities are observed, related by mirror symmetry. \textbf{f}, ARPES-measured band dispersion along the loop $P$, as marked in (c). Loop $P$ encloses the termination point of the measured Fermi arc and reveals a single left-moving chiral mode (arrow). This result shows that loop $P$ encloses a projected chiral charge $-1$ in the surface BZ, which corresponds directly to Chern number $n=-1$ in the bulk BZ. \textbf{g}, Band dispersion along loop $M$. The observed right-moving chiral mode shows an enclosed Chern number $n=+1$. \textbf{h}, Stack of MDCs along loop $M$ at different binding energies, with Lorentzian fits. Green dashed line: guide to the eye tracking the peaks. \textbf{i}, Calculated energy dispersion along $\textit{k}_{y}=-0.25\pi/a$ with the result from our ARPES spectra overlaid (open green circles).}
\label{Fig3}
\end{figure}

%Figure 3, part 1
To further explore the topological arc candidate we search for direct spectroscopic signatures of a chiral charge in PrAlGe, taking advantage of the bulk-boundary correspondence between bulk Weyl fermions and surface Fermi arcs (see Methods) \cite{Wan, arcDetect1}. We study chiral modes along straight and loop energy-momentum cuts, Fig.~\ref{Fig3}\textbf{a}, \textbf{b}. A horizontal $\textbf{\textit{k}}$ cut at $k_{y}\sim-0.25\,\pi/a$ passes through a pair of `U' states, Fig.~\ref{Fig3}\textbf{c}, \textbf{d}. We observe signatures of a left-moving and right-moving mode related by mirror symmetry. A second derivative plot of the dispersion map further confirms the observed modes and suggests additional neighboring bulk bands which approach $E_\textrm{F}$ and then turn back towards deeper binding energies, Fig.~\ref{Fig3}\textbf{e}. A comparison of this spectrum with the locations of the predicted Weyl fermions suggests that we can interpret the left- and right-moving modes as two chiral edge modes, associated with Chern number $\pm1$, Fig.~\ref{Fig3}\textbf{b}. In this way, the $\textbf{\textit{k}}$ cut is associated with a 2D momentum-space slice carrying Chern number $n_{tot}=0$, since $n_{l}=-1$ and $n_{r}=+1$ (see Methods). This again suggests that the left- and right-moving modes giving rise to two mirror-parterned `U' states are topological Fermi arcs.

%Figure 3, part 2
To provide further evidence of chiral charge in PrAlGe, we next perform an analysis of edge modes along closed loops in the surface BZ (black dashed circles labelled $P$ and $M$, Fig.~\ref{Fig3}\textbf{c}). By counting chiral edge modes on these circular paths, we search for evidence of $n_{tot} \neq 0$. Unrolling the energy-momentum dispersion for loop $P$, we observe one left-moving chiral mode that is dispersing towards $E_\textrm{F}$, Fig.~\ref{Fig3}\textbf{f}. This result unambiguously shows chiral charge $-1$ on the associated bulk manifold (see Methods). Analogously, along loop $M$ we observe a right-moving chiral mode dispersing towards $E_\textrm{F}$, Fig.~\ref{Fig3}\textbf{g}. By Lorentzian fitting of the MDCs along loop $M$ at varying binding energies, we again observe a right-moving mode, Fig.~\ref{Fig3}\textbf{h}, suggesting that the corresponding bulk manifold encloses chiral charge $+1$. As an additional check, the $\textit{ab initio}$ band dispersion calculation along $k_{y}=-0.25\,\pi/a$ shows a right-moving chiral mode dispersing towards $E_\textrm{F}$, Fig.~\ref{Fig3}\textbf{i}. An overlay of the Lorentzian fits of the chiral mode on the calculated band dispersion shows a match between our results. In this way, our ARPES spectra directly resolve topological arcs and demonstrate chiral charge through the bulk-boundary correspondence in PrAlGe.

\begin{figure}
\includegraphics[width=165mm]{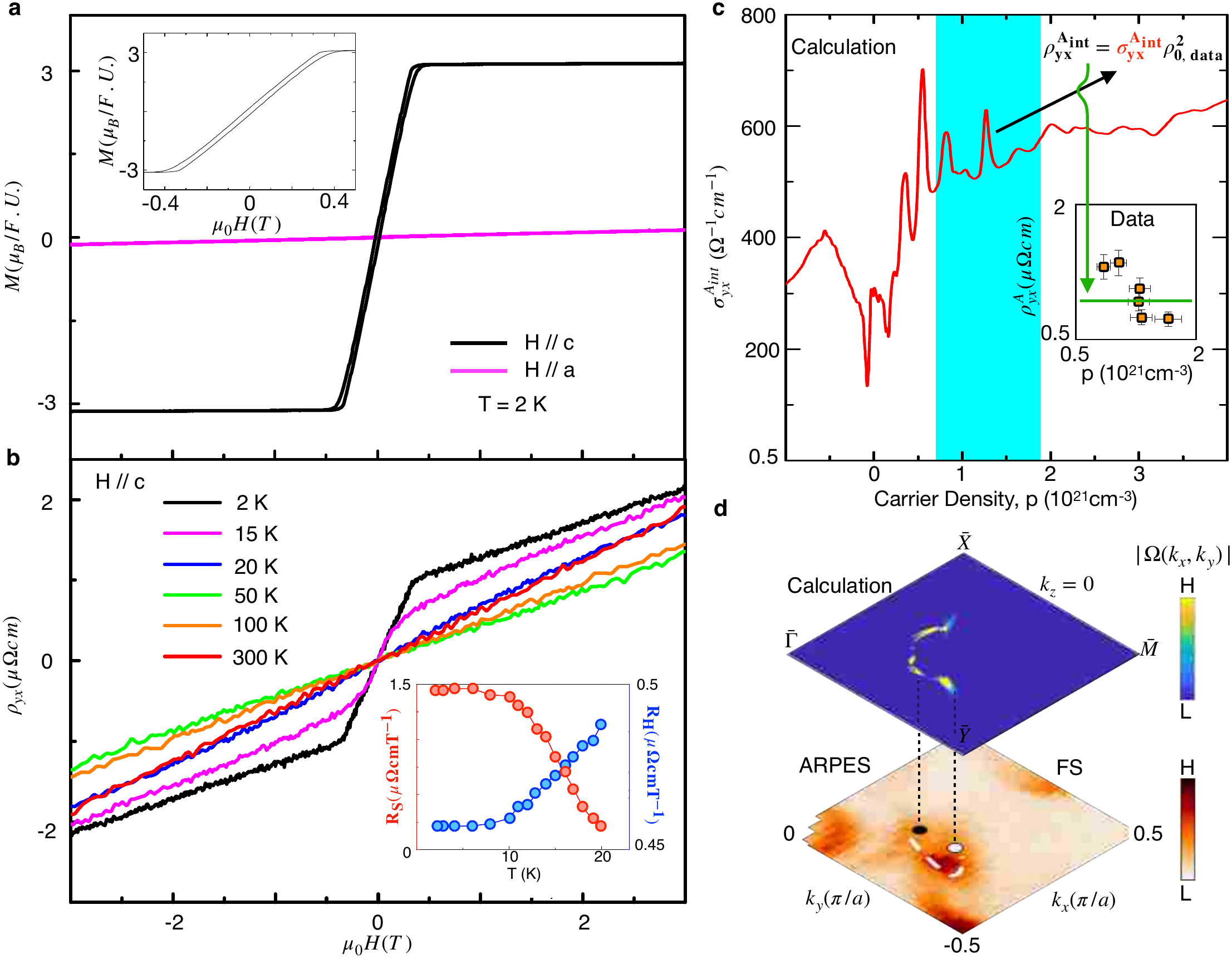}
\caption{\textbf{Intrinsic anomalous Hall response in PrAlGe.}
\textbf{a}, Magnetization $M$ of PrAlGe along the $c$-axis (black) and $a$-axis (magenta) versus magnetic induction $\mu_0$H. The Curie temperature is observed to be $T_\textrm{C} = 16$ K. Inset: zoom-in showing a hysteresis loop. \textbf{b}, Hall resistivity $\rho_{yx}$ as a function of $\mu_0$H. The inset shows the ordinary and anomalous Hall coefficients R$_H$ (blue) and R$_S$ (red) as a function of temperature extracted from the data. \textbf{c}, \textit{Ab initio} calculated intrinsic anomalous Hall conductivity $\sigma_{yx}^{A_{int}}$ as a function of carrier density. The shaded turquoise area corresponds to the carrier density of the measured crystals. The inset shows the measured anomalous Hall resistivity $\rho_{yx}^A$ as a function of carrier density $p$. The horizontal green line corresponds to the calculated intrinsic contribution to $\rho_{yx}^A$. The error bars are set by the $\pm10\%$ error in measurement of the sample dimension. \textbf{d}, Top: Berry curvature magnitude $|\Omega(k)|$ at $k_z=0$ summed over energies below the Fermi level, from \textit{ab initio} calculation. Bottom: ARPES-measured Fermi surface, suggesting that the Berry curvature field is concentrated near the termination points of the topological arc observed in ARPES.}
\label{Fig4}
\end{figure}

%Figure 4
Having demonstrated a chiral charge, we next investigate phenomena mediated by Berry curvature in PrAlGe using magneto-transport. We study the magnetization M as a function of magnetic induction $\mu_0$H (Fig.~\ref{Fig4}\textbf{a}) and observe that PrAlGe is a soft ferromagnet with the easy axis along the $c$-direction. Furthermore, we find that the transverse resistivity $\rho_{yx}$ exhibits an anomalous Hall effect, described by $\rho_{yx}$$=R_{H}$B$+\mu_{0}R_{S}$M, where $R_{H}$ is the ordinary (Lorentz-force) Hall coefficient and $R_{S}$ is the anomalous Hall coefficient \cite{AHEReview}. As shown in the inset of Fig.~\ref{Fig4}\textbf{b}, $R_{S}$ (zero for high $T$) grows rapidly towards large values while the small value for $R_{H}$ (almost invariant for different $T$) decreases very quickly below $T_c$. The observed behavior for $R_{H}$ and $R_{S}$ suggests that the anomalous Hall effect arises near $T_c$. The measured $R_{S}$ coefficient reaches a saturation value of about $1.5\,\mu\Omega\,$cm$\,T^{-1}$ at $2K$, where it dominates the response. To elucidate the origin of the observed behavior for the anomalous Hall effect, we plotted the anomalous Hall contribution $\rho_{yx}^{A}$ as a function of carrier concentration $p=1/eR_{H}$. The result shows clustered values in the vicinity of $1\,\mu\Omega$cm for different crystals, see Fig.~\ref{Fig4}\textbf{c} inset. To better understand the origin of $\rho_{yx}^{A}$, we calculated the Berry curvature contribution to the anomalous Hall conductivity, the so-called intrinsic anomalous Hall conductivity, $\sigma_{yx}^{A_{int}}$ as a function of carrier concentration, Fig.~\ref{Fig4}\textbf{c}. The calculation predicts a roughly carrier concentration-independent value of approximately $600\,\Omega^{-1}$cm$^{-1}$ with carrier densities from $p=0.9$ to $1.7\times10^{21}$cm$^{-3}$. This corresponds to an intrinsic contribution to the anomalous Hall resistivity of $\rho_{yx}^{A_{int}}=\sigma_{yx}^{A_{int}}\rho_{0}^{2} \sim 0.6\mu\Omega$cm, which we plot as a horizontal green line in inset, Fig.~\ref{Fig4}\textbf{c}. We find a remarkable agreement with the measured $\rho_{yx}^{A} \sim 1\mu\Omega$cm. This agreement suggests that the Berry curvature dominates the anomalous Hall response in PrAlGe.

%Berry curvature hot spot plot with the ab initio and ARPES measured Fermi surface
To reveal the origin of the Berry curvature fields which give rise to the intrinsic anomalous Hall response, we compare the ARPES-measured Fermi surface with the calculated Berry curvature field. By summing over energies below $E_\textrm{F}$, the Berry curvature magnitude $|\Omega(k)|$ at $k_z=0$ shows concentrated hot spots, see Fig.~\ref{Fig4}\textbf{d}. Moreover, the points of concentrated Berry curvature correspond to the position of the W$_3$ and W$_4$ Weyl fermions. A comparison with the ARPES-measured Fermi surface shows that within our momentum-space resolution the hot spot region coincides with the termination points of the measured topological arc. This suggests that the measured intrinsic anomalous Hall response arises from Berry curvature associated with the magnetic Weyl fermions \cite{Burkov2}. Our magneto-transport results provide evidence for an intrinsic anomalous Hall effect arising from Weyl fermions in the ferromagnetic phase of non-centrosymmetric PrAlGe.

 %Discussion
The surface state dispersions in our low-energy ARPES spectra and magneto-transport measurements, taken together with the topological bulk-boundary correspondence established in theory, demonstrate that PrAlGe exhibits novel Berry curvature mediated topological electronic phenomena. Our results open new research directions in understanding and engineering tunable topological electronic properties in the non-centrosymmetric magnet PrAlGe. In particular, based on our crucial investigation of the non-centrosymmetric magnetic Weyl fermion ground state, temperature-dependent ARPES measurements with further optimized energy resolution can be carried out as a next step to explore the thermally-induced phase transition and the associated Berry curvature evolution. In addition, due to the lack of time-reversal symmetry, the W$_3$ and W$_4$ Weyl fermions are offset in energy relative to one another by approximately $30$meV \cite{RAlGe_Theory}, enabling the chiral magnetic effect to be studied in devices. Furthermore, due to the absence of time-reversal and inversion symmetry, exotic types of photogalvanic effects may emerge \cite{chiral_photogalvanic, add9,add10,add11}. Topological currents in PrAlGe may also allow for the development of all-electrical spin generation and injection with no entropy production. Lastly, the soft-ferromagnetism may allow the spin-polarized topological currents to be turned on/off via an external magnetic field \cite{Fe3Sn2_Jiaxin,add12} or by varying the temperature. The unprecedented level of control and access to a large number of exotic phenomena makes the non-centrosymmetric magnet PrAlGe an exciting material platform for experimentally probing topological and quantum matter physics.

\clearpage

\clearpage

\clearpage

\end{document}